\documentclass[preprint,pra,showpacs,aps,amsmath,amssymb]{revtex4}

\usepackage{graphicx}

\begin{document}
\title{On wireless connection between Josephson qubits}

\author{Sergei Sergeenkov$^{1}$ and Giacomo Rotoli$^{2}$}
\affiliation{$^{1}$Departamento de F\'isica, Universidade Federal
da Para\'iba, 58051-970 Jo\~ao Pessoa, PB, Brazil\\
$^{2}$CNISM and DIMEG, Universit\`{a} di L'Aquila, Localit\'{a}
Monteluco, I-67040 L'Aquila, Italy \\}


\begin{abstract}
By attributing a circulating Josephson current induced diamagnetic
moment to a SQUID-type three-level qubit, a wireless connection
between such qubits is proposed based only on dipole-dipole
interaction between their moments. The estimates of the model
parameters suggest quite an optimistic possibility to
experimentally realize the suggested coupling scheme.
\end{abstract}


\maketitle

Josephson qubit is essentially a superconducting ring interrupted
by typically two or three Josephson junctions forming the basis of
an effective two level quantum
system~\cite{1a,w,1b,1c,2a,2b,2c,2d}. Most of the recently
suggested sign and magnitude tunable couplers between two
superconducting flux qubits are based on either direct or indirect
inductive coupling mediated by the SQUID~\cite{2b,2c,2d} (for
detailed and up-to-date discussion of different qubit
implementations and modern physical coupling schemes, see a
comprehensive review article by Wendin and Shumeiko ~\cite{w}).
More precisely~\cite{w}, in inductive coupling scheme a magnetic
flux induced by one qubit threads the loop of another qubit thus
changing the effective external flux. This leads to an effective
coupling between two-level qubits $H_{int}=\lambda (R)\sigma
_1^z\sigma _2^z$ with interaction $\lambda (R)$ dependent on the
length of the coupler (and thus qubit separation) $R$ via the
mutual inductance $L_{12}(R)\propto R\log R$ as $\lambda
(R)\propto L_{12}^{-1}(R)$.

\begin{figure}
\centerline{\includegraphics[width=0.85cm,angle=90]{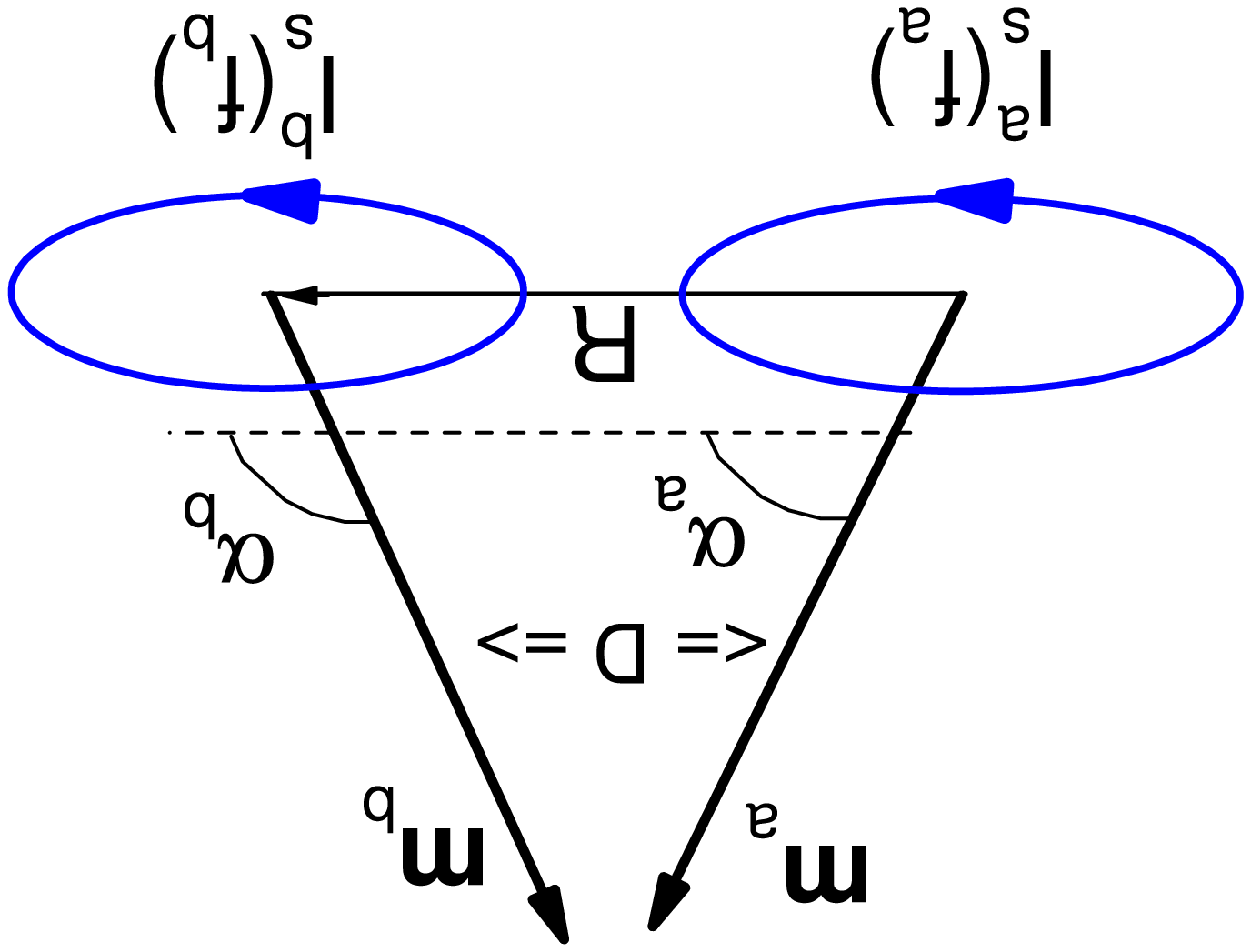}}
\vspace{1.50cm}
 \caption{(Color online) Sketch of a dipolar coupler (with strength $D$)
 between currents $I_s^{a,b}$ induced  magnetic moments ${\bf m} _{a,b}$
 of two flux qubits separated by a distance $R$. } \label{fig:fig1}
\end{figure}

In this report, we propose a wireless coupling between two
superconducting Josephson qubits based on dipole-dipole
interaction (DDI) $D$ between their diamagnetic moments. Such a
dipolar coupler has much in common with the above-mentioned
inductive coupling scheme, except that instead of the mutual
inductance controlled flux-flux interaction, we have a Zeeman-type
proximity-like magnetic interaction of a circulating current
$I_s^a$ induced dipole moment $m_a$ of one qubit with an effective
magnetic field $B_{b}$ produced by a dipole moment $m_b$ of the
second qubit, and vise versa. Thus, the principal difference of
the suggested dipolar coupler from the other coupling schemes is
the absence of electric circuit elements (like inductance and/or
capacitance) which are known~\cite{w} to be the main source of
noise and decoherence (dephasing). In this regard, the dipolar
coupler is expected to be more "quiet" than its conventional
counterparts. Besides, due to the vector nature of the DDI, the
sign of such a coupler is defined by the mutual orientation of the
qubits while its magnitude varies with the inter-qubit distance as
$D(R)\propto 1/R^3$ (see Fig.~\ref{fig:fig1}). At the same time,
as we shall show, unlike the inductive coupling scheme, the
suggested here dipolar coupler requires a three-level qubit
configuration for its implementation.

Let us consider a system of two superconducting qubits assuming,
for simplicity, that each qubit is a two-contact SQUID with a
circulating Josephson current $I_s=I_{c1}\sin \phi _1 + I_{c2}\sin
\phi _2$, where $I_{c1,2}$ is the corresponding critical current,
and $\phi _{1}$ ($\phi _{2}$) stands for the phase difference
through the first (second) contact. In turn, the circulating in
each qubit supercurrent $I_s$ creates the corresponding non-zero
diamagnetic moment ${\bf m} =I_s{\bf S}$ with ${\bf S}$ being the
(oriented) SQUID area. Recall~\cite{1a} that the quantization
condition for the total flux $\Phi ={\bf B}{\bf S} +LI_s$ (created
by the applied magnetic field ${\bf B}$ and loop self-inductance
$L$ contributions) in each SQUID is given by $\phi _1-\phi _2+2\pi
\frac{\Phi}{\Phi _0}=2\pi n$ with $n=0,1,2,..$. By introducing a
new phase difference $\theta $:
\begin{equation}
\phi _1=\theta +2\pi n, \qquad \phi _2=\theta +\frac{2\pi
\Phi}{\Phi _0}
\end{equation}
we obtain
\begin{equation}
I_s=I_{c1}\sin \theta +I_{c2}\sin (\theta +2\pi f)
\end{equation}
and
\begin{equation}
{\cal H}_s=-J_{1}\cos \theta -J_{2}\cos (\theta +2\pi f)
\end{equation}
for the circulating current and tunneling Josephson energy for
each SQUID-based qubit, respectively. Here, $f= \Phi /\Phi _0$ and
$J=\Phi _0I_c/2\pi$. In what follows, we neglect the
self-inductance of each SQUID, assuming that $LI_s\ll {\bf B}{\bf
S}$, and consider $f={\bf B}{\bf S}/\Phi _0$ as a field-induced
frustration parameter. In fact, this condition is rather well met
in realistic flux qubits~\cite{2b,2c,2d} with the so-called
degeneracy point $f=0.5$ and SQUID parameter $\beta _L=2\pi
LI_c/\Phi _0 \simeq 0.1$.

For generality, let us consider two non-identical qubits (a and b)
which are assumed to be coupled only via the DDI between their
magnetic moments ${\bf m} _{q}=I_s^{q}S_{q}{\bf \hat {e}}_{q}$
\begin{equation}
{\cal H}_{d}=\frac{\mu _0}{4\pi R^3}\left[ {\bf m}_a {\bf m}_a-
\frac{3\left({\bf m}_a{\bf R}\right)\left({\bf m}_b{\bf
R}\right)}{R^2}\right ]
\end{equation}
Here $I_s^{q}=I_{c1}^{q}\sin \theta _q+I_{c2}^{q}\sin (\theta _q
+2\pi f_q)$ is the circulating current in $q$-th qubit (in what
follows, $q=\{a,b\}$), ${\bf R}$ is the distance between qubits,
and ${\bf \hat {e}}_{q}$ is the unit vector.

Thus, the total Hamiltonian of the two coupled qubits ${\cal
H}_{tot}=\sum_q{\cal H}_s^{q}+{\cal H}_d$ reads
\begin{equation}
{\cal H}_{tot} =-\sum_{q=a,b}\left [J_{1}^{q}\cos \theta
_q+J_{2}^{q}\cos (\theta _q+2\pi f_q)\right ]+ D(f_a,f_b)\sin
\theta _a\sin \theta _b
\end{equation}
where
\begin{equation}
D(f_a,f_b)=D_1\left( 1+j_a\cos 2\pi f_a + j_b\cos 2\pi f_b +j_a
j_b\cos 2\pi f_a \cos 2\pi f_b\right )
\end{equation}
with
\begin{equation}
D_1=D_0\left( 2\cos \alpha _a\cos \alpha _b-\sin \alpha _a\sin
\alpha _b \right )
\end{equation}
Here $D_0=(J_1^{a}/2\pi )(R_0/R)^3$ with $R_0=\sqrt[3]{4\pi^2 \mu
_0J_1^{b}S_aS_b/\Phi _0^2}$ being a characteristic distance
between qubits, $j_q\equiv J_{2}^{q}/J_{1}^{q}$, and $\alpha
_{a,b}$ are the angles of ${\bf m}_{a,b}$ relative to the distance
${\bf R}$ between qubits. A sketch of the proposed dipolar coupler
is shown in Fig.~\ref{fig:fig1}. Notice that, due to its vector
character, the DDI naturally provides a sign-dependent coupling
between the qubits which is quite similar to conventional SQUID
inductance mediated coupling~\cite{2b,2c,2d}. It is also
interesting to mention that, in view of Eq.(5), DDI automatically
results in a non-conventional current-phase relation~\cite{ss}
$I_{d}(\theta _a)= \partial {\cal H}_{d}/\partial \theta _a\propto
D\cos \theta _a\sin \theta _b$ (with a sign-changeable amplitude
$D$) usually observed in $SFS$ structures and attributed to the
formation of $\pi$-type contacts~\cite{3a,3b}.

A careful analysis of the structure of the total Hamiltonian
${\cal H}_{tot}$ reveals that the inter-qubit dipole coupling $D$
introduces the transitions (mixing) between more distant states,
namely $|0>$ and $|2>$ suggesting thus that implementation of DDI
requires a three-level qubit configuration~\cite{4a,4b,4c}
(instead of its more traditional two-level
counterpart~\cite{1a,w,1b,1c,2a,2b,2c,2d}). The resulting
three-level system can be readily cast into the following form of
the qubit Hamiltonian
\begin{equation}
{\cal H}_{Q}=-\sum_{q=a,b}\left (\epsilon _qM_q^z+\Delta
_qM_q^x\right)+D\left (M_a^zM_b^z+2M_a^yM_b^y\right)
\end{equation}
where $\epsilon _q=(J_1^{q}+J_2^{q})\sin (\delta _q)\simeq
(J_1^{q}+J_2^{q})\delta _q$ and $\Delta _q=(J_1^{q}+J_2^{q})\cos (
\delta _q) \simeq (J_1^{q}+J_2^{q})$ are the energy bias and
tunneling splitting for the $q$-th qubit. Here $\delta_q\equiv
2\pi(f_q-\frac{1}{2})\ll 1$.

\begin{displaymath}
M_q^x= \frac{1}{\sqrt{2}}\left (\begin{array}{ccc}
0 & 1 & 0 \\
1 & 0 & 1\\
0 & 1 & 0
\end{array} \right ),
M_q^y=\frac{1}{\sqrt{2}}\left (\begin{array}{ccc}
0 & -i & 0 \\
i & 0 & -i\\
0 & i & 0
\end{array} \right ),
M_q^z= \left (\begin{array}{ccc}
1 & 0 & 0 \\
0 & 0 & 0\\
0 & 0 & -1
\end{array} \right )
\end{displaymath}
are the spin-$1$ analog of the Pauli matrices $\sigma_q^{\alpha}$.

\begin{figure}
\centerline{\includegraphics[width=11.0cm]{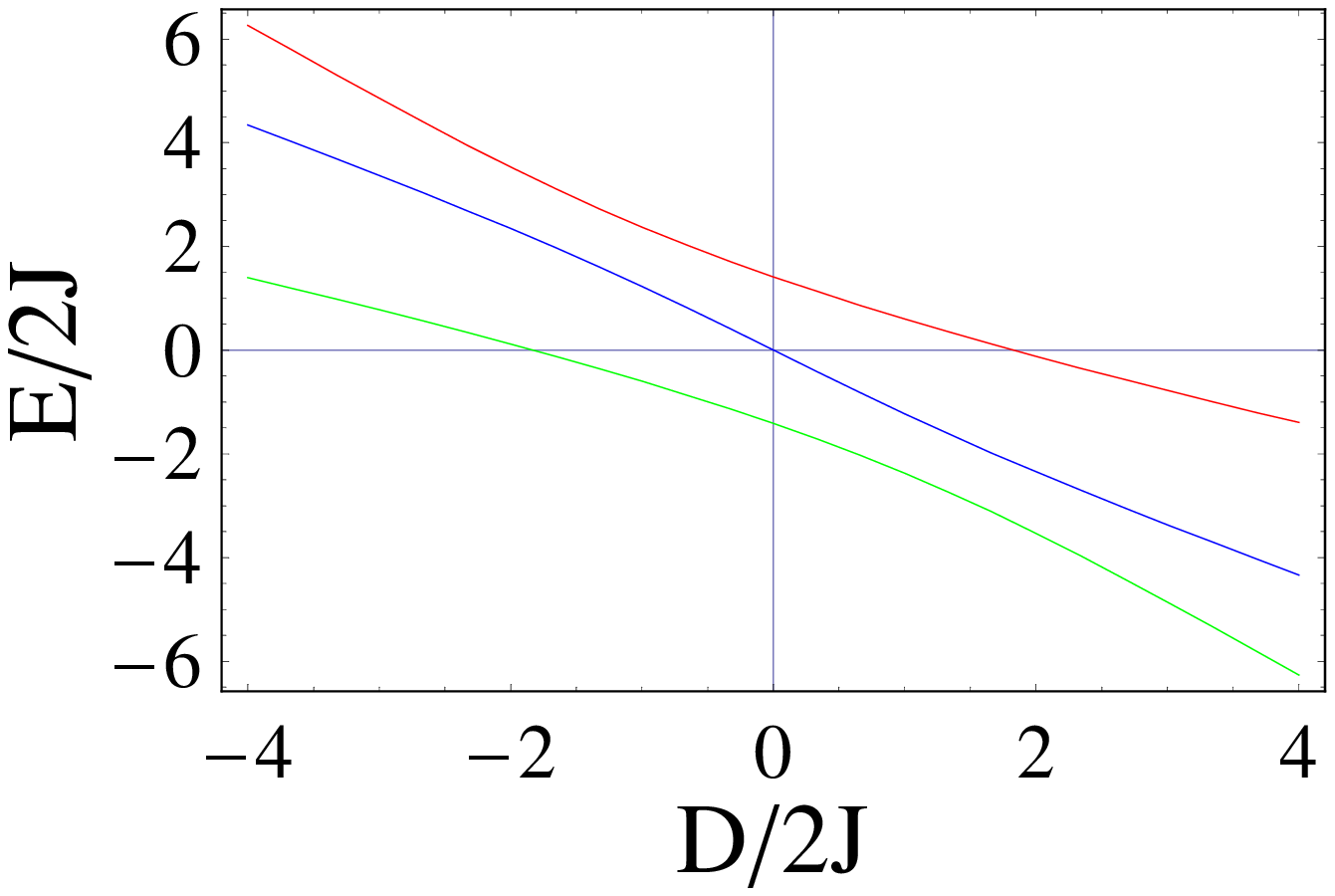}}
\vspace{0.03cm}
 \caption{(Color online) The dependence of the normalized energy levels
 $E_1$, $E_2$ and $E_3$ (from bottom to top) on
 dipole-dipole coupling $u(0,0)=D/2J$ for $f_a=f_b=1$.} \label{fig:fig2}
\end{figure}

It is easy to verify that in the absence of coupling (when $D=0$)
each individual qubit has three energy levels: $E_1^{0}=0$ and
$E_{2,3}^{0}=\pm \sqrt{\epsilon _q^2+\Delta _q^2}\simeq \pm \Delta
_q(1+\frac{1}{2}\delta_i^2)$, in accordance with the $M_q^z$
structure.

At the same time, the energy levels of the coupled qubits can be
found by solving the eigenvalue problem ${\cal H}_{Q}\Psi =E\Psi$.
As a result, we obtain the following cubic equation on the energy
spectrum $E$ of the problem (as a function of two controlling
parameters, $D$ and $f_q$)
\begin{equation}
2\Gamma ^3-\left(4\Delta^2+\frac{1}{2}D^2\right)\Gamma
-D\Delta^2=0
\end{equation}
where $\Gamma =D-E$ and $\Delta =\sum_q\Delta _q$.
\begin{figure}
\centerline{\includegraphics[width=9.5cm]{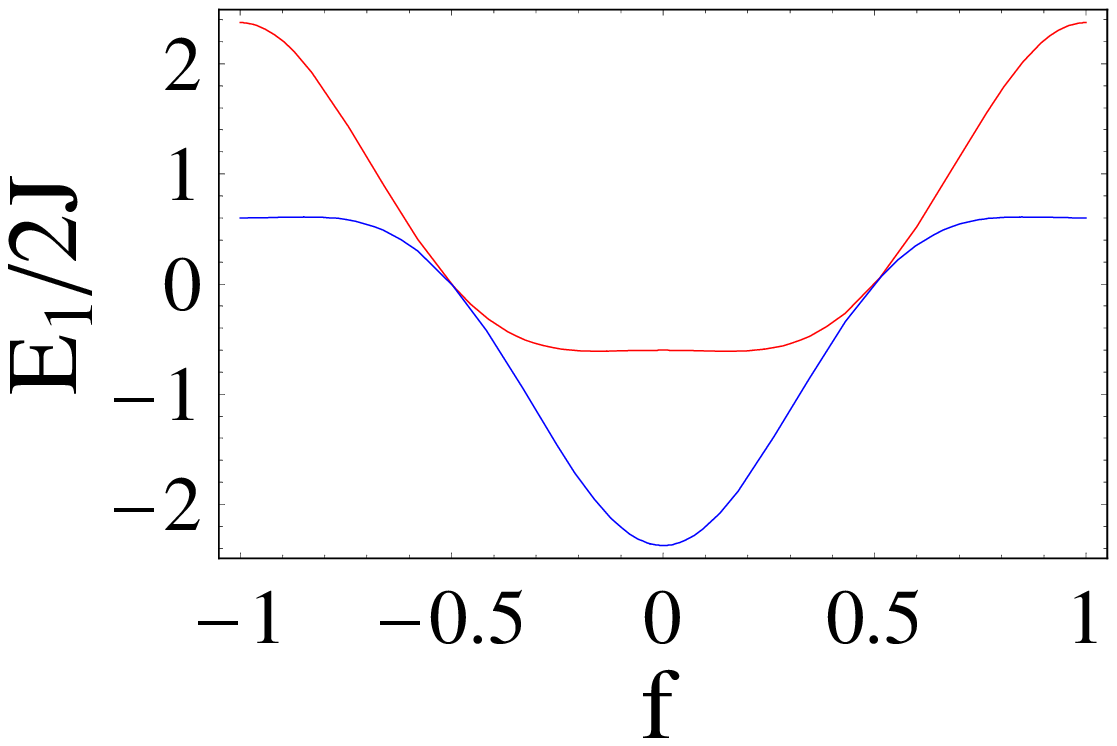}}
\centerline{\includegraphics[width=9.5cm]{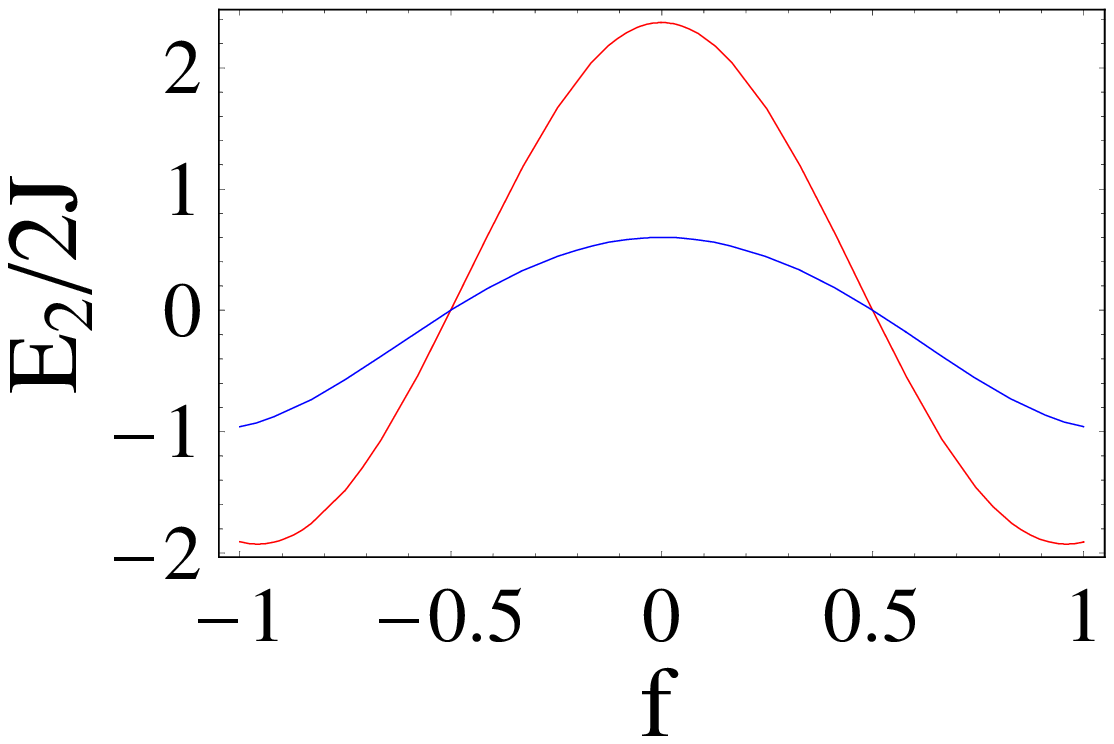}}
\centerline{\includegraphics[width=9.5cm]{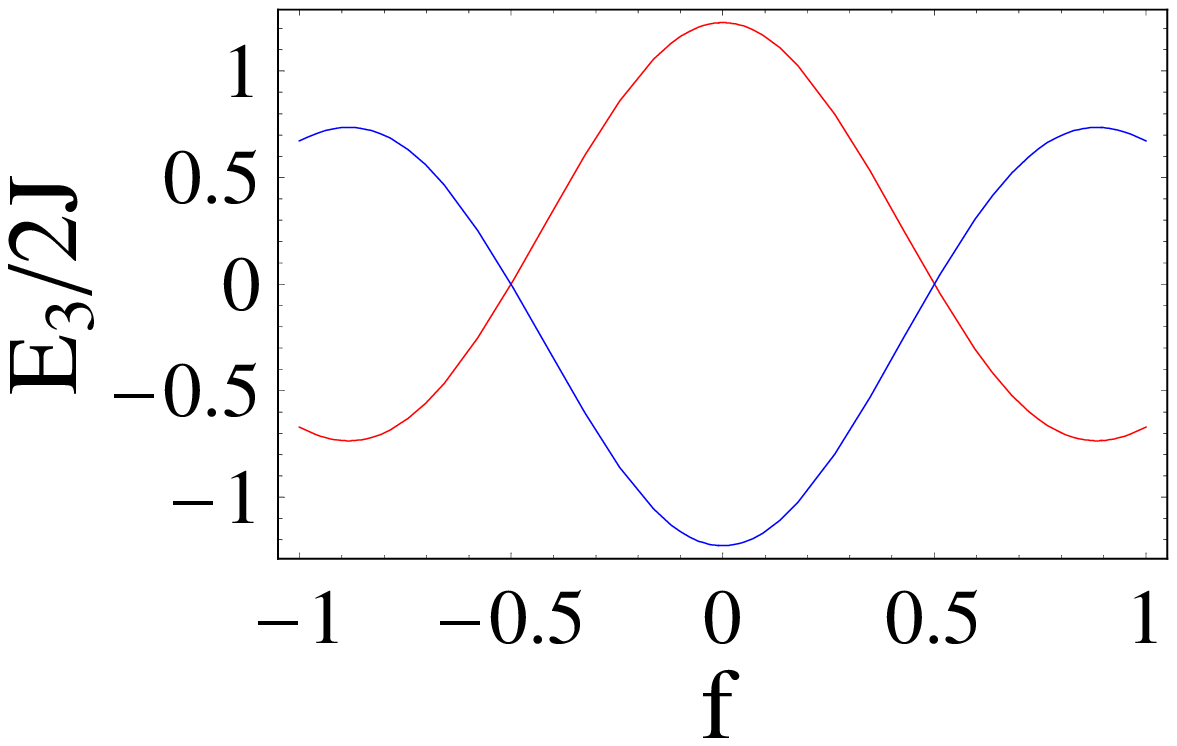}}
\vspace{0.03cm}
 \caption{(Color online) The dependence of the normalized energy levels
 $E_{1,2,3}/2J$ on frustration parameter $f_a=f_b\equiv f$ for two values
 of the dipole-dipole coupling: $u(0,0)=1$ (bottom) and $u(0,0)=-1$ (top).} \label{fig:fig3}
\end{figure}
It can be directly verified that the above cubic equation has the
following three independent solutions $E_{1,2,3}$ (corresponding
to the three-level qubit configuration), namely
\begin{eqnarray}
&&E_1=\Delta\left[u-\left(a_{+}+a_{-}\right)\right] \\ \nonumber
&&E_{2,3}=\Delta\left[u+\frac{\left(a_{+}+a_{-}\right)}{2}\pm
i\sqrt{3}\frac{\left(a_{+}-a_{-}\right)}{2}\right]
\end{eqnarray}
where
\begin{equation}
a_{\pm}=\sqrt[3]{\frac{u}{4}\pm
\frac{1}{4}\sqrt{u^2-\frac{1}{4}\left(\frac{8}{3}+\frac{u^2}{3}\right)^3}}
\end{equation}
and $u(f_a,f_b)=D(f_a,f_b)/\Delta$.

Without losing generality, in what follows we assume that
$J_1^{q}=J_2^{q}\equiv J$ (which means that $j_q=
J_{2}^{q}/J_{1}^{q}=1$ and $\Delta =2J$) and that $f_a=f_b\equiv
f$. Fig.~\ref{fig:fig2} shows the dependence of the normalized
energy levels $E/2J$ on the dipole-dipole coupling $u(0,0)=D/2J$
for $f_a=f_b=1$. Notice that $u(0,0)$ can assume negative values
due to vector nature of the dipole interaction $D$. In turn,
Fig.~\ref{fig:fig3} depicts the evolution of the coupled
three-level qubits with applied magnetic field (frustration
parameter $f$) for two different qubits orientations (given by
$u(0,0)=1$ and $u(0,0)=-1$, respectively). As would be expected
for flux qubits~\cite{1a,w}, the degeneracy point is situated near
$f=0.5$.
\begin{figure}
\centerline{\includegraphics[width=14cm]{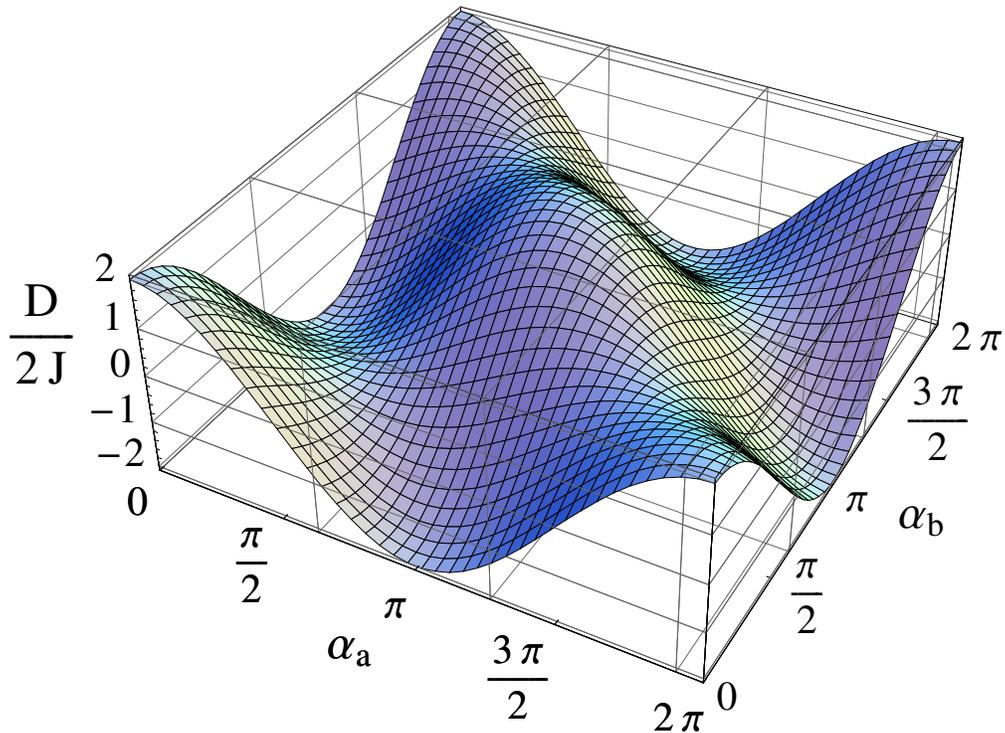}}
\vspace{0.03cm} \caption{(Color online) The dependence of the
dipolar coupler strength on angles between qubits shown in
Fig.~\ref{fig:fig1}}. \label{fig:fig4}
\end{figure}
The angular dependence of the DDI amplitude $D$ on $\alpha_{a,b}$
is shown in Fig.~\ref{fig:fig4}. Notice that, like in inductive
based coupling scheme~\cite{w,2b,2c,2d}, the dipolar coupler may
change its sign from positive (when two moments are parallel to
each other) to negative (for the anti-parallel configuration) or
even disappear (when two moments are perpendicular to each other).
In turn, Fig.~\ref{fig:fig5} depicts variation of
$D_1(\alpha_q,R)$ as a function of the normalized distance $R/R_0$
between qubits for three values of $\alpha $ (assuming the
parallel orientation with $\alpha _a=\alpha _b=\alpha$, see
Fig.~\ref{fig:fig1}).
\begin{figure}
\centerline{\includegraphics[width=11cm]{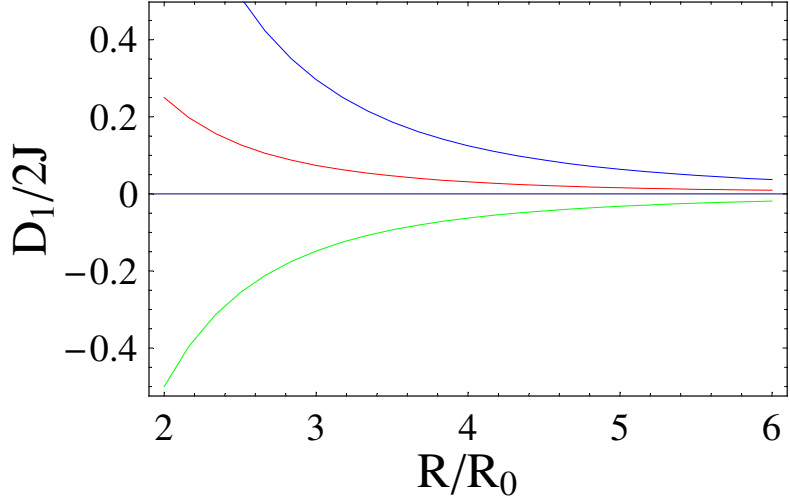}}
\vspace{0.03cm}
 \caption{(Color online) The dependence of $D_1(\alpha_q,R)$ on the
 distance $R$ between qubits for three values
 of $\alpha $ (from top to bottom: $\alpha =\frac{\pi}{2}$, $\frac{\pi}{3}$,
 and $\frac{\pi}{4}$).} \label{fig:fig5}
\end{figure}
Finally, let us estimate the main model parameters based on the
available experimental data on long-range couplers.
Using~\cite{2c} $S=50\mu m \times 50\mu m$ and $I_c=0.5\mu A$ for
the area of a single qubit and SQUID critical current, we obtain
$R_0=\sqrt[3]{2\pi \mu _0I_cS^2/\Phi _0}\simeq 100\mu m$ for DDI
characteristic separation which results in the following estimate
for DDI mediated coupler frequency (see Fig.~\ref{fig:fig5})
$\Omega=D/h=(J/2\pi h)(R_0/R)^3\simeq 1GHz$ for the inter-qubit
distance of $R=400\mu m\simeq 4R_0$. This value remarkably
correlates with the frequencies achieved in the mutual inductance
mediated couplers~\cite{w,2b,2c,2d}, suggesting quite an
optimistic possibility to experimentally realize the proposed here
dipolar coupling scheme.

In summary, a theoretical possibility of wireless connection
between Josephson qubits  (based on dipole-dipole interaction
between their induced magnetic moments) was proposed and its
experimental realization was briefly discussed.

This work was supported by the Brazilian Agency CAPES and the
Italian MIUR PRIN 2006 under the project \textit{Macroscopic
Quantum Systems - Fundamental Aspects and Applications of
Non-conventional Josephson Structures}.

\end{document}